\documentclass{article}

\usepackage{arxiv}

\usepackage[utf8]{inputenc} 
\usepackage[T1]{fontenc}    
\usepackage{hyperref}       
\usepackage{url}            
\usepackage{booktabs}       
\usepackage{amsfonts}       
\usepackage{nicefrac}       
\usepackage{microtype}      
\usepackage{lipsum}
\usepackage{graphicx}

\title{Synchronization of complex human networks}

\author{
  Shir Shahal, Ateret Wurzberg, Inbar Sibony, Hamootal Duadi \\
   Faculty of Engineering and the Institute of Nanotechnology and Advanced Materials\\
  Bar-Ilan University\\
  Ramat Gan 5290002, Israel \\
  \texttt{} \\
   \And
 Elad Shniderman, Daniel Weymouth \\
  Department of Music\\
  Stony Brook University\\
  Stony Brook, NY 11794, USA \\
  \texttt{} \\
   \AND
   Nir Davidson \\
   Department of Physics of Complex Systems \\
   Weizmann Institute of Science \\
   Rehovot, Israel\\
   \texttt{} \\
   \And
   Moti Fridman \\
    Faculty of Engineering and the Institute of Nanotechnology and Advanced Materials\\
  Bar-Ilan University\\
  Ramat Gan 5290002, Israel \\
  \texttt{mordechai.fridman@biu.ac.il} \\
}

\begin{document}
\maketitle

\begin{abstract}
The synchronization of human networks is essential for our civilization, and understanding the motivations, behavior, and basic parameters that govern the dynamics of human networks is important in many aspects of our lives. Human ensembles have been investigated in recent years, but with very limited control over the network parameters and in noisy environments. In particular, research has focused predominantly on all-to-all coupling, whereas current social networks and human interactions are often based on complex coupling configurations, such as nearest-neighbor coupling and small-world networks. Because the synchronization of any ensemble is governed by its network parameters, studying different types of human networks while controlling the coupling and the delay is essential for understanding the dynamics of different types of human networks. We studied the synchronization between professional violin players in complex networks with full control over the network connectivity, coupling strength of each connection, and delay. We found that the usual models for coupled networks, such as the Kuramoto model, cannot be applied to human networks. We found that the players can change their periodicity by a factor of three to find a stable solution to the coupled network, or they can delete connections by ignoring frustrating signals. These additional degrees of freedom enable new strategies and yield better solutions than are possible within current models. Our results may influence numerous fields, including traffic management, epidemic control, and stock market dynamics.
\end{abstract}

\keywords{Complex networks \and Synchronization \and Human networks}

The synchronization of coupled ensembles appears in numerous fields, including biology~\cite{sumpter2006principles,conradt2009group}, astronomy~\cite{strogatz2004sync}, psychology~\cite{wasserman1994social,morris2005social}, optics~\cite{roy1992dynamical,deshazer2001detecting,fridman2010passive}, economics~\cite{saavedra2011synchronicity} and politics; at different size scales, from the synchronization of planets~\cite{strogatz2004sync} to the synchronization of subatomic particles~\cite{strogatz2012sync}; and in different time-scales, from slow-moving mechanical structures~\cite{strogatz2005theoretical,arane2009coupling} to coupled ultrafast lasers~\cite{schibli2003attosecond,fridman2012measuring}. Synchronization is crucial for the life of all living species on our planet~\cite{sumpter2006principles,conradt2009group}, from the cellular level~\cite{davis2001biological,oleskin2014network} to the crowd synchrony of large groups~\cite{buhl2006disorder}. In particular, the synchronization of human networks is essential for our civilization~\cite{javarone2017evolutionary,werner2007dynamics,krause2010swarm} and can impact the physical and mental well-being of individuals in groups~\cite{wasserman1994social,morris2005social}. Understanding the motivations, behavior, and basic parameters that govern the dynamics of human networks is important for many aspects of our lives, including stock market dynamics~\cite{saavedra2011synchronicity}, traffic management~\cite{li2007relationship}, epidemic control~\cite{porfiri2008synchronization} and investigating the decision-making processes in different types of groups~\cite{sumpter2012six,smaldino2012origins,conradt2005consensus,sueur2016social}. Additionally, studying the dynamics of human networks will help predict the consequences of introducing artificial intelligence into our highly connected world, where each node in a computer network will have complex decision-making ability~\cite{russell2016artificial,krogh1995neural}.

Human ensembles and crowd synchrony have been investigated in recent years. Synchronized brokers in the stoke market were found to earn more money~\cite{saavedra2011synchronicity}, the synchronization of crowd attention was shown to be a basic survival mechanism~\cite{gallup2012visual,sun2017perceiving}, pedestrians walking on the London Millennium bridge synchronized their footsteps through the bridge vibrations to form macroscopic oscillations of the bridge above a critical number~\cite{strogatz2005theoretical}, the collective movement of concert audiences showed vortexes and gaslike states~\cite{silverberg2013collective,mendez2018density}, the synchronized movements of dancers differ from those of nondancers~\cite{miura2011coordination, boker2005synchronization}, and an audience clapping hands shows both synchronization and period doubling~\cite{neda2000self,neda2000physics}.

However, all these seminal studies had limited control over the network parameters, namely, the connectivity of the network, coupling strength, and delay between individuals, and were subject to noisy environments. In particular, these studies focused mostly on all-to-all coupling, whereas current social networks and human interactions are often based on complex coupling configurations. To date, there are no studies of the synchronization of complex human networks, e.g., one-dimensional, two-dimensional, scale-free or small-world connectivity~\cite{strogatz2001exploring,strogatz2018nonlinear}. Additionally, the influence of changing the coupling strength or the delay between two individuals is critical for the dynamics of the network and has not been studied in human networks thus far.

\begin{figure}[htbp]
\centering
\includegraphics[width=\textwidth]{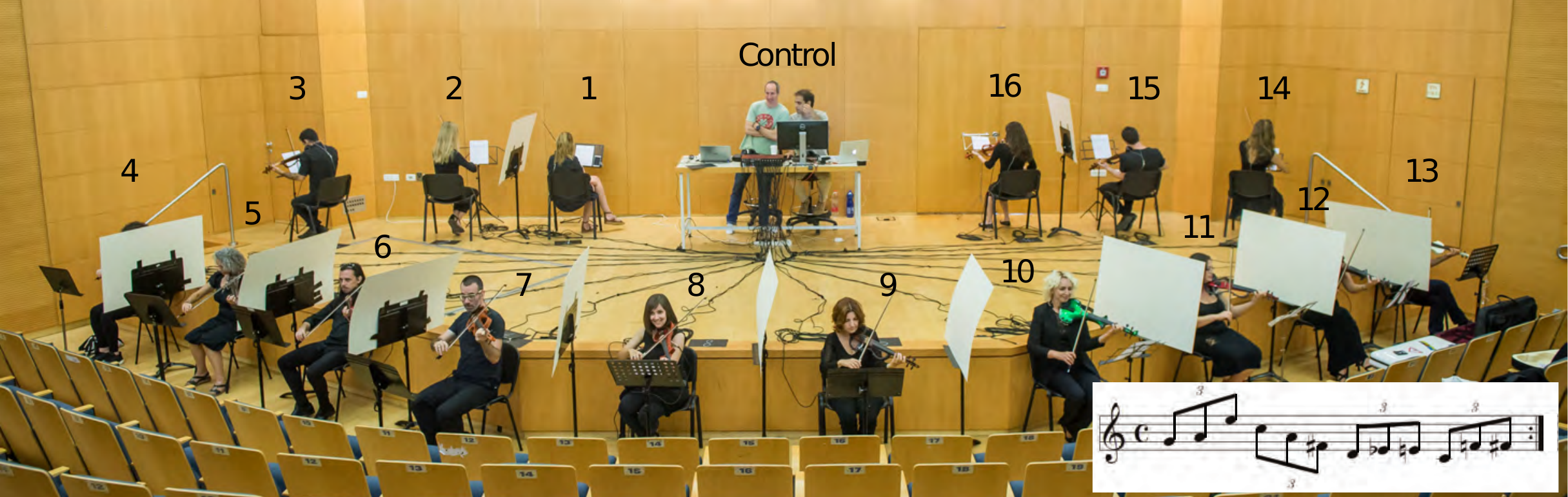}
\caption{Sixteen coupled electric violin players repeating a musical phrase presented in the inset. The audio output from each violin is connected to our computer-controlled mixing system. Then, the mixing system sends to the headphones of each player a sum of audio signals of the desired connectivity, strength and delay.}
\label{fig:system}
\end{figure}

We studied the synchronization between professional violin players in complex human networks with full control over the network connectivity, the coupling strength of each connection, and the delay. We set sixteen isolated electric violin players to repeatedly play a musical phrase. We collected the output from each violin and controlled the input to each player via noise cancellation headphones. The players could not see or hear each other apart from what was heard in their headphones. All the players started playing together with the help of an external conductor who set the rhythm during the first period. The only instruction to the players was to try synchronize their rhythm to what they hear in their headphones. A picture of the experimental setup is shown in Fig.~\ref{fig:system}, and the musical phrase is shown in the inset. We established different network connectivities and introduced delayed coupling between the players while monitoring the phase, periodicity, volume and frequency of each player with a mixing system.~\footnote{The measured raw data is available online: www.eng.biu.ac.il/fridmama/research/data/}

Our results reveal that the usual models for coupled networks such as the Kuramoto model~\cite{kuramoto1987statistical,kuramoto2012chemical,strogatz2000kuramoto,acebron2005kuramoto} cannot be applied to human networks. We found that the players can change their periodicity to find a stable solution to the coupled network with delayed coupling~\cite{neda2000self,neda2000physics,taylor2010spontaneous} and deleting connections by completely ignoring frustrating signals. These additional degrees of freedom enable new strategies and yield better solutions than are possible within the simple Kuramoto model. To analyze the dynamics of a human network and the influence of different parameters on its global behavior, we developed a new model based on the Kuramoto model that takes into account these important abilities of the human mind, which have been neglected thus far. A detailed description of the model is presented in the supplementary materials.

\begin{figure}[htbp]
\centering
\includegraphics[width=\linewidth]{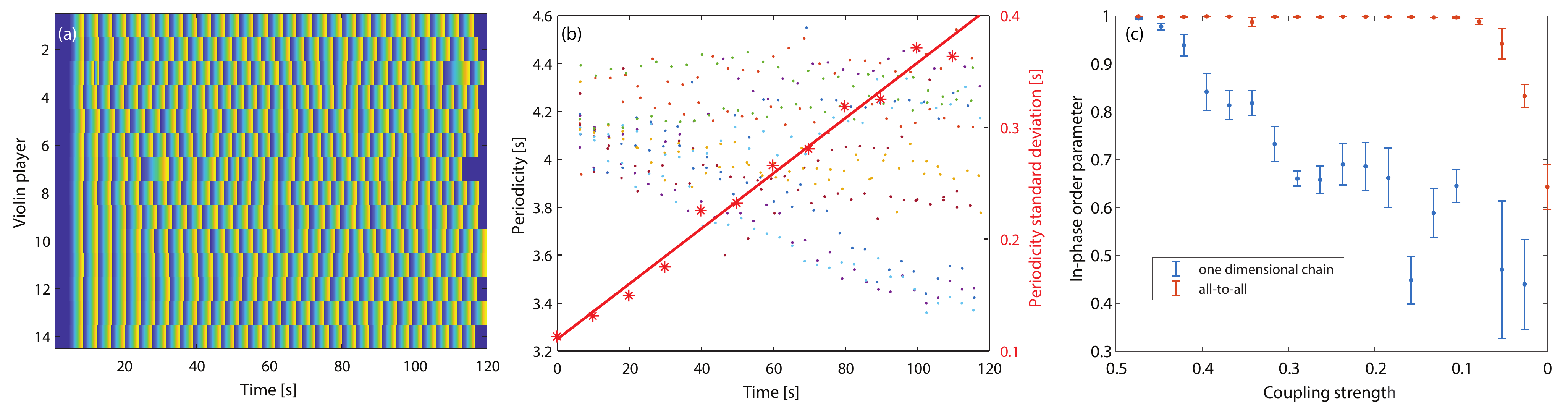}
\caption{(a) The phase of each violin player as a function of time without coupling, where blue denotes the beginning of the musical sentence and yellow denotes the end. (b) The periodicity of each player (color dots) and the standard deviation of the periodicity (asterisks) with a fitted linear curve as a function of time when the coupling is set to zero, showing that with no coupling, each player is changing its periodicity without any correlation to other players. (c) In-phase order parameter of the network as a function of the coupling strength for two different configurations: one-dimensional chain with nearest neighbor coupling and a global all-to-all coupling. The error bars are obtained by repeating the same configuration several times.}
\label{fig:coupling}
\end{figure}

In our first experiment, we set the coupling between the players to zero, causing the players to hear only themselves. We measured the time it took for each of the players to play the musical phrase and denoted this time as the periodicity of the player. In Fig.~\ref{fig:coupling}(a), we show the phase of each player as a function of time, where blue denotes the beginning of the musical phrase and yellow denotes the end. In Fig.~\ref{fig:coupling}(b), we show the periodicity of all the players and the standard deviation of the periodicity as a function of time. The opening phrase, accompanied by an external rhythmical beat, verified that all the players started with the same periodicity; after the beat stopped, the periodicity of each player deviated towards his natural one. The periodicities of the players were spread as a function of time, reflecting that the players could not hear each other.

Then, we introduced coupling between the different players with our mixing system. The coupling strength between each pair of players was calibrated by a logarithmic scale of the volume~\cite{stevens1965loudness}. We arranged the players in two configurations, a one-dimensional open chain and an all-to-all coupling where each player was coupled to all the others. In each configuration, we started with a coupling strength of 0.5 and reduced it linearly to zero over a period of 4 minutes. We measured the in-phase order parameter in the network as a function of the coupling strength and present the results in Fig.~\ref{fig:coupling}(c). The in-phase order parameter is calculated by $\left< \cos \left( \varphi_i - \varphi_j \right) \right>$, where $\varphi_i$ is the phase of the $i$'th player, $\varphi_j$ is the phase of its coupled neighbor, and we average over all connections. Similar to other networks, the order parameter of the all-to-all configurations remained high for lower coupling strength than the one-dimensional configuration. \footnote{The order parameter does not reach zero since the playing time is limited to 4 minutes to keep the players focused.}

\begin{figure}[htbp]
\centering
\includegraphics[width=8cm]{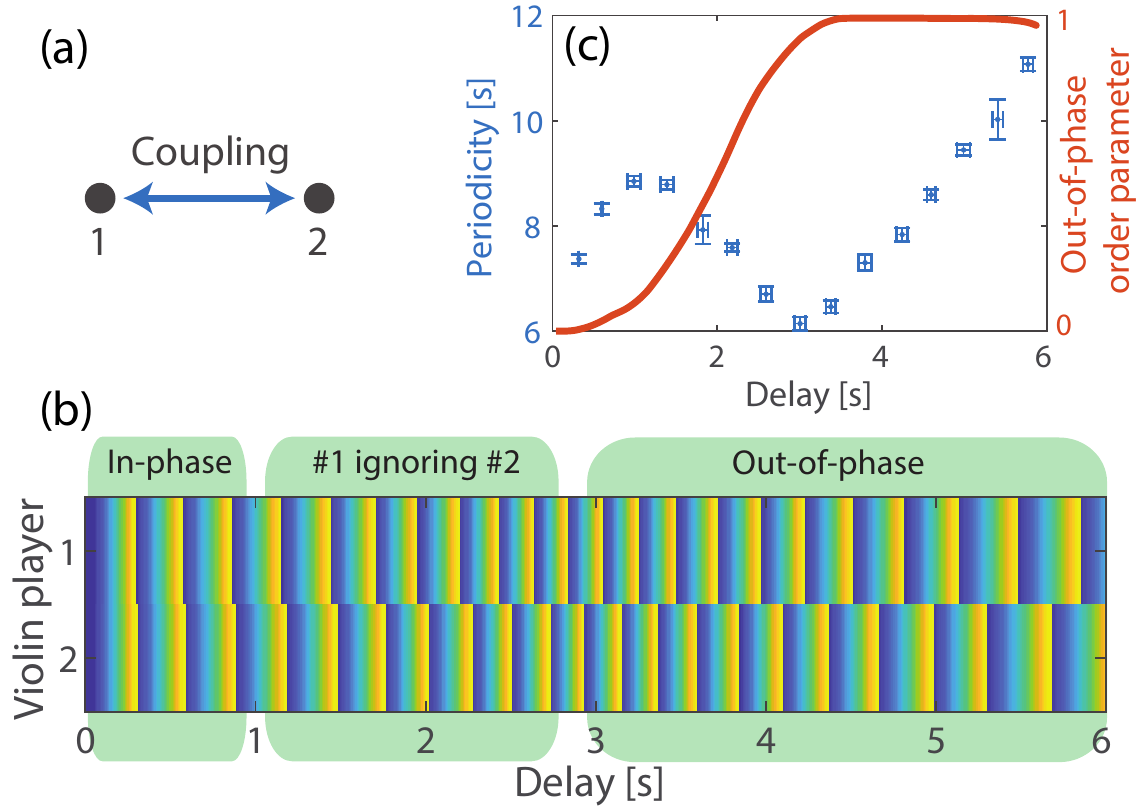}
\caption{Two coupled violin players with a delay between them, as schematically shown in (a). (b) The phase of each player along the musical phrase. We emphasize three states of synchronization: in phase; player \#1 is ignoring player \#2; out of phase. (c) The periodicity of the players together with the out-of-phase order parameters as a function of the delay averaged over a moving window.}
\label{fig:MD2}
\end{figure}

Next, we set the coupling strength to 0.5, which is strong enough to insure synchronization, as shown by Fig.~\ref{fig:coupling}(c). Then, we imposed a delay on the coupling between the players, starting from zero delay and slowly increasing it until the delay is equal to the duration of the musical phrase. The delay prevents the players from synchronizing with each other, which leads them to shift from in-phase synchronization to other states of synchronization. We demonstrate these states of synchronization by examining the synchronization of two coupled violin players as a function of the delay, schematically shown in Fig.~\ref{fig:MD2}(a). In Fig.~\ref{fig:MD2}(b), we present the progress of each player by a color code. In Fig,~\ref{fig:MD2}(c), we show the periodicity as a function of the delay together with the out-of-phase order parameter, $\left< \sin \left(\varphi_i - \varphi_j \right) \right>$. The results reveal three states of synchronization, namely:
\begin{itemize}
\item Initially, the delay is zero, so the two players are perfectly synchronized in phase. With the introduction of the delay, they increase the periodicity (play slower) to keep the delay small relative to the duration of each note. This state is emphasized on the left side of Fig.~\ref{fig:MD2}(b) and is indicated by the increased periodicity, presented in Fig.~\ref{fig:MD2}(c). This effect was also observed when playing over the Internet with a small delay~\cite{chafe2010effect}.
\item When the delay was further increased, the players could not maintain an in-phase synchronization state, as one of them started to ignore the other and returned to its original periodicity. In our case, player \#1 ignored player \#2, while player \#2 still followed player \#1, which is emphasized in the middle part of Fig.~\ref{fig:MD2}(b).
\item When the delay was increased to approximately half of the periodicity, an out-of-phase synchronization emerged that satisfied both players. This state is highly stable; therefore, when the delay is further increased, the players increase the periodicity to ensure that the periodicity is always double the delay. This is shown in Fig.~\ref{fig:MD2}(c), where the out-of-phase order parameter is presented by the red curve. Once this order parameter approaches unity, it stays there, and the periodicity increases linearly with the delay. This is also observed by the checkerboard pattern on the right side of Fig.~\ref{fig:MD2}(b).
\end{itemize}

\begin{figure}[htbp]
\centering
\includegraphics[width=\linewidth]{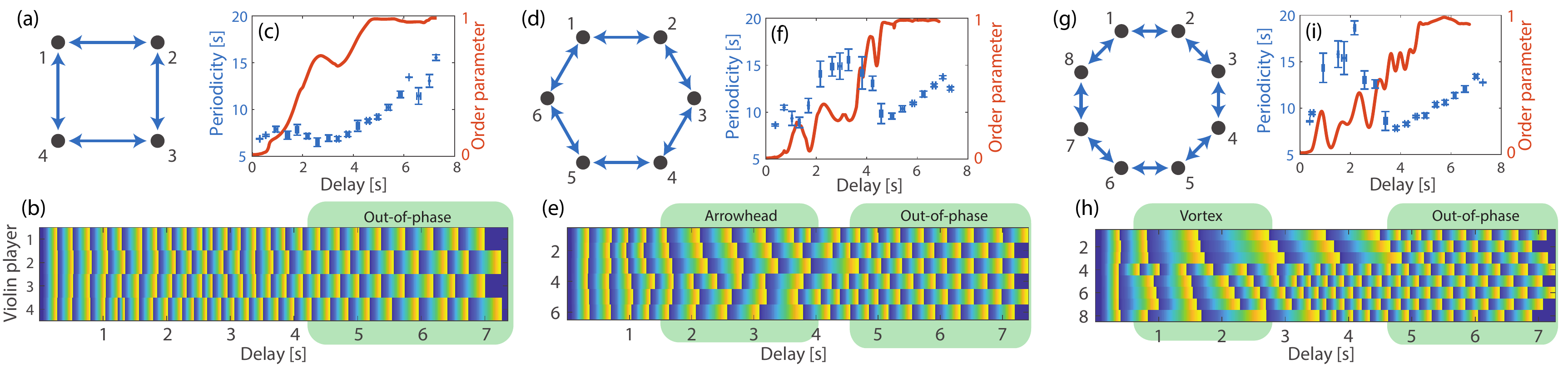}
\caption{Four, six, and eight violin players coupled as schematically shown in (a), (d) and (g) with increasing delay between the players. The phase of each violin player along the musical phrase as a function of the delay is shown in (b), (e) and (h). When the delay is low, an in-phase synchronization is observed; when the delay is increased, we see a vortex or an arrowhead synchronization; and when the delay is half of the periodicity, we see a stable out-of-phase synchronization. In (c), (f) and (i), we present the periodicity as a function of the delay together with the out-of-phase order parameter. As shown, when the players experience out-of-phase synchronization, indicated by an order parameter of unity, the periodicity increases linearly with the delay, remaining twice the delay to preserve the highly stable out-of-phase state of synchronization.}
\label{fig:MD4to8}
\end{figure}

When increasing the number of the coupled violin players to 4, 6 or 8, as shown in Fig.~\ref{fig:MD4to8}(a), (d) and (g), they follow the same behavior as the delay is increased: we first observe an in-phase synchronization with a reduction in periodicity; next, each player spontaneously decides to ignore one of its inputs. If all the players are ignoring the same side, they create a vortex synchronization, as seen in Fig.~\ref{fig:MD4to8}(h), while if some are choosing one side and others choose another side, they create an arrowhead-shaped synchronization, as seen in Fig.~\ref{fig:MD4to8}(e). Finally, when the delay reaches approximately half of the periodicity, a stable and highly ordered out-of-phase synchronization emerged, as evidenced by the checkerboard pattern emphasized at the right side of Fig.~\ref{fig:MD4to8}(b), (e) and (h), together with the linear increase in periodicity as a function of the delay, as seen in Fig.~\ref{fig:MD4to8}(c), (f) and (i) and the out-of-phase order parameter, which approaches unity.

\begin{figure}[htbp]
\centering
\includegraphics[width=\linewidth]{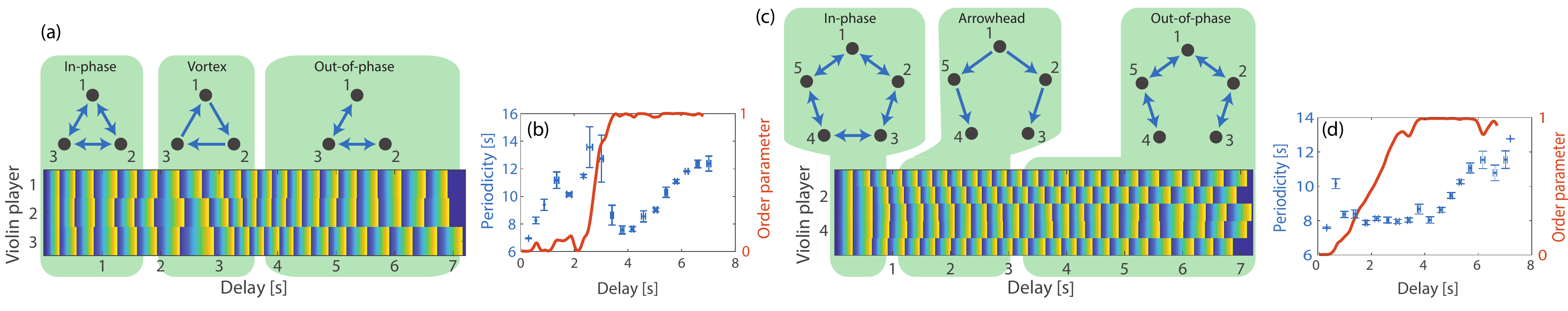}
\caption{Three and five coupled violin players as an example of an odd number of coupled players where the out-of-phase state is no longer stable. In these cases, the players change the connectivity and ignore one of the links, reducing the system to an open chain where out-of-phase synchronization is stable again. (a) Three coupled players showing in-phase, vortex, and out-of-phase synchronization states. (b) The periodicity and the open-chain out-of-phase order parameter as a function of the delay, showing that once the players obtain an out-of-phase synchronization state, they maintain it by increasing the periodicity with the delay. (c) Five coupled players, showing in-phase, arrowhead, and out-of-phase synchronization states. (d) The periodicity and the open-chain out-of-phase order parameter, showing the same behavior as the three players.}
\label{fig:MD3to5}
\end{figure}

For odd numbers of violin players, the out-of-phase state of synchronization is no longer stable~\cite{nixon2011synchronized,d2008synchronization,takamatsu2001spatiotemporal}. In such cases, the players spontaneously choose to ignore one of the connections, which breaks the chain to form an open chain where the out-of-phase synchronization state is possible. Thus, the players change the connectivity of the configuration into one with a stable solution. In Fig.~\ref{fig:MD3to5}, we present the results for three and five coupled violin players. When the delay is low, the players remain in an in-phase synchronization, as shown on the left side of (a) and (c), while increasing the periodicity, as shown in (b) and (d). When we increase the delay, the players choose either a vortex state, as shown in (a), or an arrowhead state, as shown in (c). Finally, when the delay reaches half of the periodicity, the players prefer the out-of-phase state of synchronization while ignoring one of the connections, as shown on the right side of (a) and (c). When this state is achieved, it is highly stable, as seen by the out-of-phase order parameter shown in (b) and (d) calculated for open-chain connectivity. When the delay is further increased, the players increase the periodicity, keeping it twice the delay, to maintain the out-of-phase synchronization state, as shown in (b) and (d).

\begin{figure}[htbp]
\centering
\includegraphics[width=8cm]{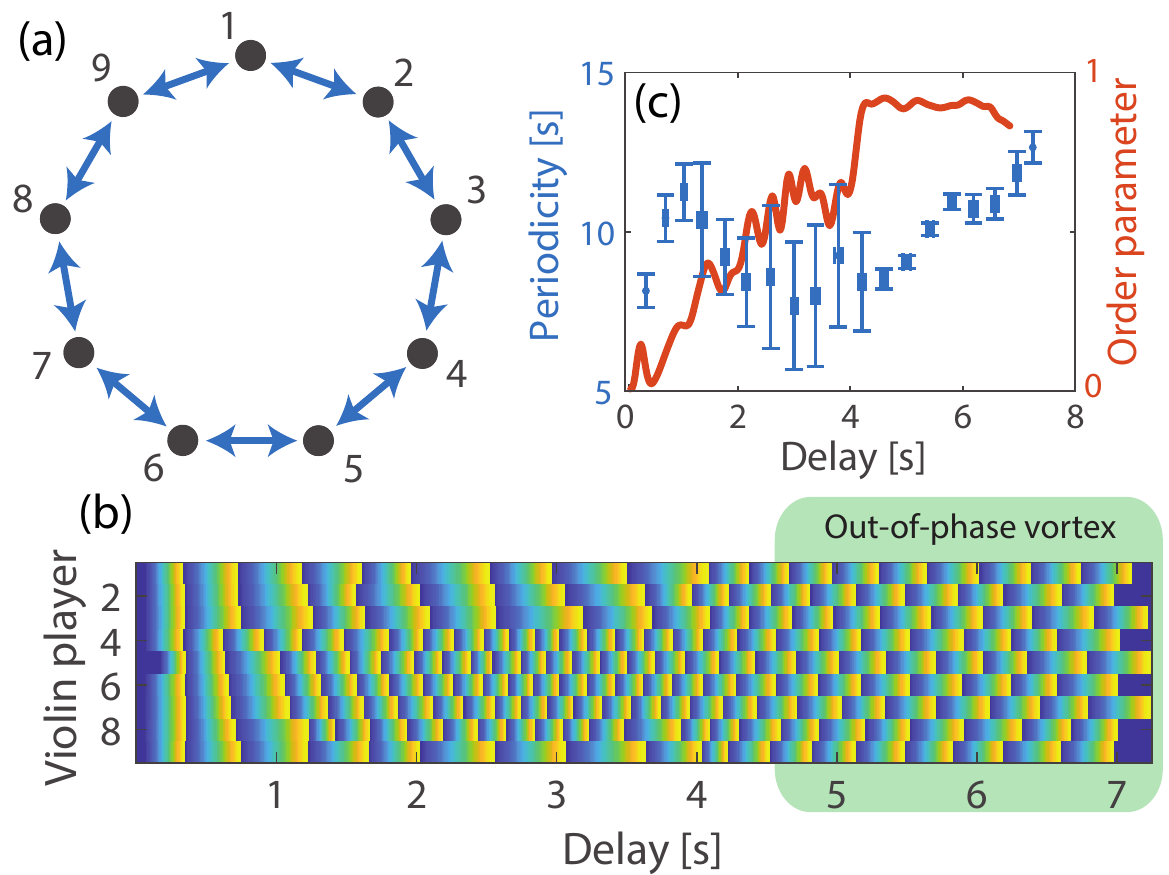}
\caption{Nine coupled violin players illustrating an out-of-phase vortex state. (a) Schematics of the coupled players. (b) The phase of each player along the musical phrase, showing the checkerboard pattern on the right, indicating an out-of-phase state; (c) periodicity and out-of-phase order parameter as functions of the delay. The order parameter reaches 0.9 due to the vortex but stays there while increasing the periodicity as a function of the delay, indicating a stable state.}
\label{fig:MD9}
\end{figure}

For nine or more coupled violins, as schematically shown in Fig.~\ref{fig:MD9}(a), the violin players can find an approximate out-of-phase synchronization state without breaking the connection by shifting each player by $2 \pi / 9$ toward the top of the out-of-phase synchronization. The combination of an out-of-phase state with a vortex state is shown on the right side of Fig.~\ref{fig:MD9}(b). We evaluated the out-of-phase order parameter, which reaches 0.9 instead of unity due to this vortex shown in Fig.~\ref{fig:MD9}(c). Nevertheless, this state is as stable as the regular out-of-phase states, as evidenced by the increase in periodicity as a function of the delay while keeping the order parameter at 0.9.

\begin{figure}[htbp]
\centering
\includegraphics[width=\linewidth]{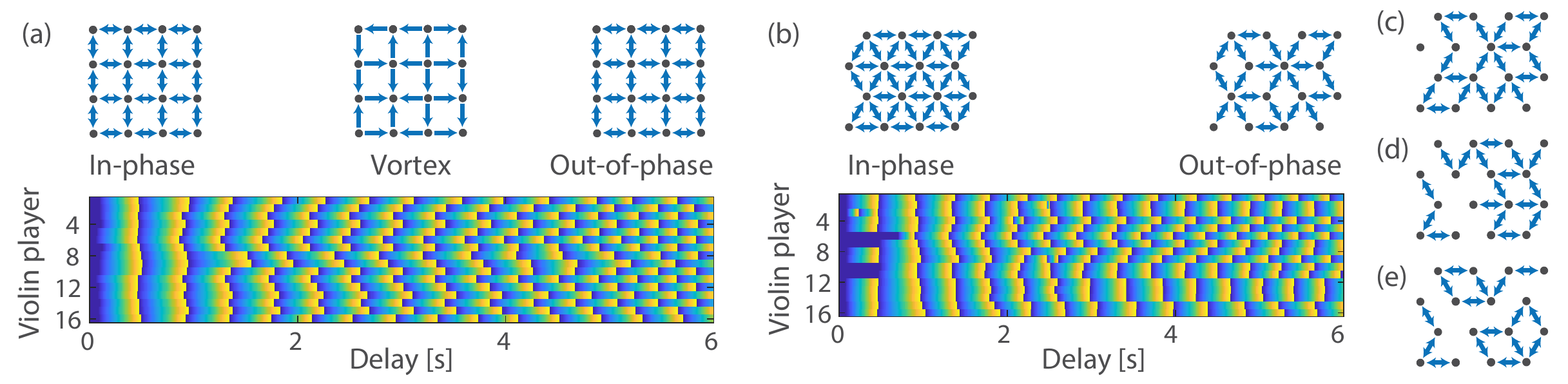}
\caption{Sixteen violin players arranged in two two-dimensional lattices: square and triangular configurations. (a) The evolution of the square lattice configuration as a function of the delay showing in-phase, vortex and out-of-phase states of synchronization. (b) The evolution of the triangular configuration showing that for the out-of-phase synchronization state, the connectivity of the network changes to obtain a stable solution. We repeated the experiment and obtained a different solution each time based on the same motifs. Three representative solutions are shown in (c) - (e).}
\label{fig:2d}
\end{figure}

Finally, we measured the synchronization of the players when arranging them in square lattice and in triangular lattice configurations while increasing the delay. The results are shown in Fig.~\ref{fig:2d}. the measured results of the square lattice are shown in Fig.~\ref{fig:2d}(a) and the measured results of the triangular lattice are shown in Fig.~\ref{fig:2d}(b). For zero delay in the square lattice configuration, the players are synchronized in phase, and with increased delay, they create vortex states until reaching the out-of-phase state of synchronization, which is a stable solution to the square configuration. In the triangular configuration, the players start with in-phase synchronization, and with increased delay, they cannot find a stable solution~\cite{nixon2013observing,nixon2011synchronized}, so they ignore some of the connections and change the connectivity of the network to one based on square motifs or open chains. This result is shown by the resulting lattice on the right side of Fig.~\ref{fig:2d}(b). When repeating the experiment, the players converge to a different solution every time, as shown in Fig.~\ref{fig:2d}(c), (d) and (e), presenting solutions that include rings of four and six players and the breaking of the network into smaller coupled clusters.

To develop a model for coupled human networks, we extended the simple Kuramoto model for coupled oscillators~\cite{kuramoto1987statistical,kuramoto2012chemical,strogatz2000kuramoto,acebron2005kuramoto} to include broad-bandwidth oscillators and the ability of each oscillator to ignore some of the connections. A detailed description of the model, including all the stages in its development, is presented in the supplementary materials. First, we simulated even numbers of coupled oscillators while increasing the delay between them with the simple Kuramoto model. We found that as the delay increases, the periodicity likewise increases until an out-of-phase state of synchronization is achieved. However, since the oscillators were narrow-band, they did not shift their periodicity by more than 15\%. Therefore, the system could not maintain the out-of-phase state of synchronization upon increasing the delay, so the system switched to a different state. Then, we switched to broad-bandwidth oscillators~\cite{wang2008stability} and repeated the same simulations. Indeed, the system found the out-of-phase state of synchronization and stayed in it by increasing the periodicity together with the delay, as observed in the experiment.

For odd numbers of players, the simple Kuramoto model failed to reproduce the measured results and showed only vortex states of synchronization~\cite{nixon2013observing,nixon2011synchronized}. Therefore, we extended the model to include the ability to delete contradicting connections. This extended model showed the same dynamics as we measured for odd numbers of coupled violin players, including the deletion of connections and reduction of the network until enabling a stable out-of-phase state of synchronization. We tried different mechanisms for choosing which connections to delete, including keeping connections with similar periodicity, keeping connections with similar phases, and choosing randomly which connections to delete and which to keep. The results show that all these mechanisms led to the same dynamics. Therefore, we conclude that the mechanism by which each player chooses which connection to delete and which to keep is not significant. As long as each player can delete a connection, the network will change the connectivity until it finds stable out-of-phase state of synchronization. Therefore, the specific psychology of each player has no role in the macroscopic network dynamics of coupled violin players.

To conclude, we investigated the synchronization of different types of complex human networks where all the parameters of the networks are under control. We measured the phase and synchronization of coupled violin players in different network configurations and when introducing delay between the coupled players. We found that human networks differ from previously studied networks in their ability to adjust the periodicity and the network connectivity by ignoring a coupled player and effectively deleting the connection. When we coupled an even number of players on a ring, the players found a stable out-of-phase synchronization state and tuned their periodicity accordingly as the delay was increased. When we coupled an odd number of players on a ring, the players changed their connectivity and then adjusted their periodicity. Our system will be extended to investigate decision making models in different configurations, bifurcation and phase transition in human networks, and the nonlinearity of crowds. This research may impact numerous fields, including economics, decision making research, epidemic spreading, information transfer modeling, traffic control, and more.

\bibliographystyle{unsrt}

\end{document}